\documentclass[12pt]{elsarticle}

\usepackage{graphicx,bbm,amssymb,amsopn}

\newcommand{\ii}{ {\rm i} }
\newcommand{\dd}{ {\rm d} }
\newcommand{\ZZ}{\mathbb{Z}}
\newcommand{\RR}{\mathbb{R}}
\newcommand{\CC}{\mathbb{C}}

\newcommand{\LL}{{\hat {\cal L}}}

\def\tr{{\,{\rm tr}\,}}

\def\one{\mathbbm{1}}

\begin{document}

\title{Observables and density matrices embedded in dual Hilbert spaces}

\author{T. Prosen$^{1,2}$, L. Martignon$^3$ and T.H. Seligman$^{4,2}$}
\address{$^1$ Departement of Mathematics and Physics, University of Ljubljana, Ljubljana, Slovenia}
\address{ $^2$ Centro Internacional de Ciencias, Cuernavaca, Morelos M\'exico}
\address{$^3$ Institute of Mathematics and Computer Science, Ludwigsburg University of Education, Ludwigsburg, Germany}
\address{ $^{4}$Instituto de Ciencias F\'\i sicas---Universidad Nacional Aut\'onoma de 
M\'exico, Cuernavaca, Morelos, M\'exico}

\begin{abstract}
The introduction of operator states and of observables in various fields of quantum physics has raised 
questions about the mathematical structures of the corresponding spaces. In the framework of third quantization 
it had been conjectured that we deal with Hilbert spaces although the mathematical background was not entirely clear, particularly, 
when dealing with bosonic operators. This in turn caused some doubts about the correct way to combine bosonic and fermionic 
operators or, in other words, regular and Grassmann variables.  In this paper we 
present a formal answer to the  problems on a simple and very general basis. We illustrate the resulting construction 
by revisiting the Bargmann transform and finding the known connection between ${\cal L}^2(\RR)$ and the Bargmann-Hilbert space.
We pursue this line of thinking one step further an discuss the representations of complex extensions of linear canonical transformations as 
isometries between dual Hilbert spaces.
 We then use the formalism to give an explicit formulation for Fock spaces involving both fermions and bosons thus solving the problem 
at the origin of our considerations.
\end{abstract}
%\pacs{03.65Aa,03.67.-a}

\maketitle

\section{Introduction}

The description of open systems, while dating back to the early days of quantum 
mechanics \cite{VNeumanbook}, has acquired increasing relevance in more recent 
years due to the burgeoning interest in quantum information processes, where 
decoherence is probably the strongest limiting element for practical implementation. 
Thus states have to be described by density operators, which can appear in many forms, 
such as density matrices or Wigner functions \cite{Schleichbook}. This made it attractive 
to consider states and operators that represent e.g. observables on 
as similar as possible a footing \cite{prosen1}. An additional important step in this direction 
was achieved when  Fock spaces for density operators and observables were introduced \cite{prosen1, 
prosen2, prosenseligman1, prosenseligman2}. The original formulation  was given for 
fermionic systems, where anti-commuting operators (i.e., Grassmann variables) 
lead to finite dimensional spaces as long as we keep the number of variables finite, because 
each variable can only appear to powers zero and 1. For bosonic systems where 
partial differential operators and/or their commutation relations are commonly used, 
arbitrary powers are allowed and thus infinite dimensional spaces occur. In 
reference \cite{prosenseligman1} it was shown that the use of the Cauchy-Schwarz 
inequality could prove the relevant expressions to be convergent. It was suggested 
that the use of dual Hilbert spaces both isomorphic to 
${\ell}^2$ would provide an adequate framework to formulate the bosonic  
case as well as the case of mixed systems \cite{prosenseligman2}. In this 
paper we shall present a very natural
framework of dual Hilbert spaces that takes care of the problems we address 
and may, potentially, have wider use. Note that in a more abstract mathematical context, dualities in Hilbert spaces of operators have been 
extensively discussed, see e.g. Ref.~\cite{pisier}, however, to best of our knowledge, the application of these ideas to quantum mechanics of 
open or statistical systems have not been abundant.

Let us consider an operator $H$ as an observable and operator $S$ as a state expressed e.g. in matrix form. Then the trace
of $H S$ represents the expectation value of the operator $H$ for the state $S$.
The bilinear form between the space containing the observables and the one containing the states
is the essential ingredient for our considerations. We have proven in previous work \cite{prosenseligman1}
that the bilinear form yields convergent results for all physically relevant situations \cite{prosenseligman1}. We also conjectured that
 both the space in which the relevant observables lie and the space in which the states lie are
separable infinite dimensional Hilbert spaces. While it seems that the first of these two results is all we need, 
it would be very useful to prove the second in order to ensure that all results obtained for bosonic and fermionic 
Fock spaces are consistent. Furthermore future applications could be treated without concerns about convergence.
We will thus use a bilinear form of the type of the trace expression for the expectation value to 
derive the Hilbert space structure of both spaces formally. As the trace mentioned is central to our standard description of open systems, 
we expect easy returns in this field, yet the construction we will present is not depending on this specific bilinear form.
Technically we shall define the two vector spaces by choosing finite or infinite dimensional sets of equal dimension of 
linearly independent vectors in each of the two vector spaces. From these we construct the pertinent Hilbert spaces by 
choosing sequences, which are limits of linear combinations, with coefficients in $l^2$.

In the following section we shall present this natural but very general, formal  approach 
and illustrate it thereafter with the Bargmann transform from ${\cal L}^2(\RR)$ to Bargmann-Hilbert space \cite{Bargmann} as
 an elegant and instructive example in section 3. We complete this section by showing that representations of complex extensions 
 of linear canonical transformations \cite{loebl} can be formalized as isometries or partial isometries between Hilbert spaces  \cite{plebanski} and
 and this reflects on the theory of "Harmonic analysis on phase space" \cite{Folland}. This subject suggest a wide scope of
 applications from nuclear physics to optics and beyond. We shall illustrate this by showing that a series of singular integral transforms used in nuclear reactions 
 since many years can be neatly fit into our framework eliminating known problems of singularities \cite{Horiuchi, NPAMS} in a unified way. Other applications will be mentioned.
 In section 4 we shall proceed to the modern application, that gave rise to this present study. We shall give a 
unified description of an open many-body system containing both fermions 
and bosons in terms of Fock spaces for both the (mixed) states and the observables. While giving a more rigorous mathematical frame
to this application is the central purpose of this paper, we shall speculate in an outlook on further 
uses of the results obtained as well as their limitations.

\section{Dual pairs of Hilbert spaces}
We start out with two vector spaces say of functions, operators or matrices and consider two countable sets  $ \{v_i, y_i; i=1,\ldots, \infty\}$, one from each vector space, 
of linearly independent elements, either finite with an equal number of elements or countably infinite. We stipulate the existence of a bilinear form or product
between elements, one from each of these sets, that fulfills
\begin{equation}
v_i \times y_j \equiv \delta_{i,j}
\label{delta}
\end{equation}
 We then define two vector spaces  $V$ and $Y$ with elements
 \begin{equation}
v= \sum_{i=1}^{\infty} \phi_i v_i \in V,
\quad
{\rm and}
\quad
y=\sum_{i=1}^{\infty} \omega_i y_i \in Y,
\end{equation} 
where the $\phi_i$ and $\omega_i$ are real or complex numbers and 
chosen such that the column vectors ${\bf \phi}$, ${\bf \omega}$ 
with components $\phi_i$, $\omega_i$ are elements of $\ell^2$ 
under the standard scalar product ${\bf \phi}^t {\bf\omega}$ 
for vectors of real or complex numbers, constituting the 
$\ell^2$~closure of the finite linear combinations.
Based on the Cauchy-Schwarz inequality the bilinear form 
\begin{equation}
v \times y = \sum_{i=1}^{\infty} \phi_i \omega_i (v_i \times  y_i) 
\label{productbasisvectors}
\end{equation} 
is well defined, and we can use it to define a scalar product and 
hence a measure on the spaces $V$ and $Y$. These spaces then become 
Hilbert spaces isomorphic to $\ell^2$ in an abstract sense.  
This scalar product is constructed by defining the    
adjoint of any vector $v \in V$  or $y \in Y$ as
\begin {equation} 
v^* = \sum_{i=1}^{\infty} \bar{\phi_i} y_i,\;\; y^* = \sum_{i=1}^{\infty} \bar{\omega_i} v_i.
\label{dualvector}  
\end{equation}

This allows us to define the scalar product in each space for 
$v,u \in V$ and $y,x \in Y$ with appropriate component vectors ${\bf \phi, \chi, \omega, \xi}$ as
\begin{equation}
v \cdot u \equiv v \times u^* = {\bf \phi}^t \bar {\chi},\quad 
y \cdot x \equiv y^* \times x ={\bar {\bf \omega}^t \xi},
\label{scalarproducts}
\end {equation}
where $^t$ indicates a transposition. 
The isomorphism $T$ from $V$ to $Y$ may be written in a simple symbolic form
\begin{equation}
T= \sum_j y_j v_j
\end{equation}
 and similarly for its inverse.  Again the Cauchy-Schwarz inequality guarantees 
 that this map and its inverse are well defined on the respective spaces.

Note however that the measures 
so defined might in general not be the most common ones used in either of the spaces.  
It is possible that the spaces $V$ and $W$ are identical also as 
function or operator spaces; in this case this construction implies 
that the $v_i$ and $w_i$ are dual basis sets in the same space.
This will particularly be true if the spaces are Fock spaces for 
fermions with a finite number of underlying states \cite{prosen1}.

\section{The Bargmann transform and the representation of complex extensions of linear canonical transformations}

Integral kernels of the form of an exponential with an exponent, which is a quadratic form in both sets of variables, have a very aide range of applications. 
In particular the unitary representations of linear canonical transformations can be formulated in this way \cite{MoshinskyQuesne, Folland}, the Fourier transform being the most common example corresponding to the exchange of coordinates and momenta. Complex extensions of
these representations \cite{loebl} have found a wide range of applications, which we will mention below, but we shall state why considering the special case of the Bargmann transform \cite{Bargmann} because of its elegance and transparency.

\subsection{The Bargmann transform}
The Bargmann transform is defined as a transformation between ${\cal L}^2({\RR)}$ and the Bargmann-Hilbert space of analytic functions over 
$\CC$. According to Bargmann (\cite{Bargmann}, Eqs.(2.1),(2.9a) and (2.10)), the
Bargmann kernel, for one dimension, is given by  

\begin{equation}  A(z, q) = \pi ^{-1/4} 
\exp\left[-\frac{1}{2}(z^2+q^2)+ \sqrt{2} zq)\right] = \sum_n \phi_n(q) z^n 
\end{equation}
Here 
\begin{equation}
\phi_m (q) = [2^m m! \sqrt{\pi}]^{-1/2} e^{-q^2} H_m(q)
\end{equation}
are the normalized Hermite functions, which form an orthonormal 
basis of ${\cal L}^2({\RR)}$. $H_m(q)$ are the Hermite polynomials (physicist's version).
The monomials
\begin{equation}
\frac{z^m}{\sqrt{m!}}
\end{equation}
are a complete orthonormal basis of the Bargmann-Hilbert space, which is defined as a space of
holomorphic functions over the complex numbers with the measure
\begin{equation}
d\mu(z) = e^{z\overline z} d{\rm Re}(z) d{\rm Im} (z) 
\end{equation}

It becomes clear, that we can 
now proceed exactly as described in the previous section. Defining 
one basis as the monomials $z^j/\sqrt{j!}$ and the other as the Hermite functions of $q$. 
we have the relation to the formal vectors of the previous section:
\begin{equation}
v_i \equiv \phi_i(x), \\ 
y_j \equiv z^j/\sqrt{j!} \\
\end{equation}
and the transformation $T$ between the spaces spanned by the two basis sets
is given as an integral transformation with the Bargmann kernel or its inverse.

The scalar products within each of the spaces as well as the transformation 
between the two spaces are given. The bilinear form on vectors of the two spaces 
is given by inverting the argument implicit in eq.(5) i.e.
\begin{equation}
v\times y = v^* \cdot y = v\cdot y^*
\end{equation}
where $v^*$ is the Bargmann transform i.e. the dual of $v$ and  $y^*$ is the inverse Bargmann transform 
i.e. the dual of $y$. Unitarity of the Bargmann transform guaranties that the last equation holds.
This implies 
\begin{equation}
v_i \times y_j = \phi_i (x) \times z^j/\sqrt{j!} = \delta_{i,j}, 
\end{equation}
and  thus we reach the starting point of the argument in the previous section.

We thus have not used the argument of the previous section, 
but rather  proceeded in the opposite direction. We used the measures in each subspace and the Bargmann transform
and show that the bilinear form we require emerges automatically.
Note the elegance that Bargmann's description brings into the scheme. If we had defined 
$z$ as real we would still get a solution, but it certainly would {\it not} be elegant. 
This indicates, that the usefulness of the technique will largely depend on the 
elegance of the solution we find for the scalar products in the individual spaces, and there we have some freedom. 
Yet the advantage of the technique will rather lie in the use of the transform, i.e. 
in the use of the dual spaces and possibly a scalar product in one of them, but 
it may well happen that the ``obvious" internal scalar product is not the one 
that results form the desired bilinear relation. Indeed, had we used the 
generating function for Hermite polynomials rather than the Bargmann transform,
orthonormality of the Hermite polynomials would be required, which implies a 
different measure on $\RR$,  that includes the Gaussian necessary for convergence.

As a final note of this subsection, consider that the Bargmann variables are closely related 
to raising operators of the Harmonic oscillator $a^\dagger$. We can then use the monomials
 $(a^\dagger)^n/\sqrt{n!}$ as a basis in a Hilbert space of operators, and use
Hermite functions as the dual set of basis vectors. The elegance of Bargmann's construction 
resides precisely in avoiding the use of these operators and defining a 
proper Hilbert space of analytic functions. Nevertheless we can see that 
our construction is also valid for operators, if an appropriate 
bilinear form between the basis sets of the two spaces is established. 

The original definition of Bargmann space and all related 
quantities is actually for many complex variables, so it allows the description of a many boson system. For fermions, anti-commuting 
creation and annihilation operators are  to physicists the most familiar way to use Grassmann 
variables. It is thus natural to use an operator picture for mixed Fermion-Boson systems and even more so, 
if  we treat open systems as we shall do in the next section. 

\subsection{Representations of complex extensions of canonical transformations}

Again we concentrate on the case of one degree of freedom for which linear canonical transformations are given as simplectic  $2\times 2$ matrices
 \begin{equation}
{\cal S} =  \left ( \begin{array} {cc} a & b \\ c & d \end{array} \right )
\end{equation}

with $a\, d - b\, c = 1$  for real matrix elements. This matrix acts on a vector of momentum $p$ and coordinate $q$. Its unitary (ray-) representation $U({\cal S})$ in coordinates $q, q^\prime$ is 
given as an integral transform with the kernel
\begin{equation} 
(2 \pi b )^{-1/2} \exp[-(i/2)(q{^\prime} ^2\, {d/b} - 2 q^\prime q /b + q^2 \, {a/b})]
\end{equation}
If the matrix elements of ${\cal S}$ are chosen complex we have a complex extension of the canonical transformation and the above kernel can  be viewed as representing this extension in the sense, that if the integral of the folding of two kernels exists and is well defined,then we obtain (up to a phase) the representation of the corresponding matrix product \cite{loebl}.
In particular the Bargmann transform would correspond to the representation of the complex extension

%\begin{alignat}{2}
%\sigma_x &= \begin{pmatrix} 0 & 1\\ 1 & 0\end{pmatrix} \; , \quad
%\sigma_y &= \begin{pmatrix} 0 & -\rmi\\ \rmi & 0\end{pmatrix} \; , \quad
%\sigma_z &= \begin{pmatrix} 1 & 0\\ 0 & -1\end{pmatrix} \; .
%\label{PauliMats}\end{alignat}

\begin{equation}
{\cal S_{\rm Bargmann}} = \left ( \begin{array} {cc} \sqrt{2} & i/ \sqrt{2}  \\ -i \sqrt{2} & \sqrt{2} \end{array} \right ).
\end{equation}

We can thus interpret the Bargmann kernel as an isometric representation of this complex extension. Note that the normalization is not relevant for these identifications when we consider maps between different spaces. It is also worthwhile mentioning that the older literature 
talks about unitary representations \cite{loebl}. This results from the fact that all infinite dimensional separable Hilbert spaces are equivalent as Hilbert spaces, but not necessarily as function spaces and we prefer to maintain this difference as in reference \cite{plebanski}. This is all the more reasonable because partial isometries have appeared frequently in this field \cite{moshvarios} to which the same argument would apply. 
It suggests itself to interpret the wide range of complex extensions that have been used in a similar manner
using the formalism of the previous section. This will allow us to proceed even if we cannot find a solution 
as elegant as the one of Bargmann for the measure in the new space.  

Let us first remark, that, whenever the kernel is a generating function for a well known basis of functions, a Bargmann space like construction is possible as is nicely illustrated by the construction of Barut and Girardello \cite{BarutG} for Laguerre polynomials, which has also found applications \cite{Zahn-BG} though certainly on a reduced scale when compared to the Bargmann transform. Yet a caveat should be taken into account; If numerical work has to be done with the kernel the spaces of functions over complex variables have also serious disadvantages, as the dimension is doubled and the analytic structure of Bargmann like spaces is not easily exploited. A second caveat pertains to the sequential application of complex extensions: if we change the space we use, as in the case of the Bargmann transform applying it twice makes no sense as the domain of the variables would not coincide. Only the application of the inverse is meaningful. If on the other hand the domain of the variable is maintained successive applications of the same of the same or different transformations are possible if the corresponding integrals exist, and thus semi-groups can be generated.

By way of example of applications of kernels that can be represented as complex extensions of linear canonical transformations in quantum problems,  we shall turn to
 the use of Gaussian kernels in nuclear reactions. This happens in the the closely related resonating group \cite{HW}  and generator coordinate \cite{BM} methods. In this framework transformations with Gaussian kernels greatly simplify the anti-symmetrization of systems with finite numbers of Fermions in a translationally invariant setting. In an early review \cite{Horiuchi} discusses the problems arising from the singularities of the transforms used, which have been analyzed in \cite{NPAMS}. The solution in terms of the Bargmann transform \cite{JPG-SZ} has been successfully applied \cite{Hecht} but the singular transforms have still been essential for numerical work. 
We shall limit our considerations to the one dimensional case as again the generalization is obvious. The kernels and the complex $2 \times 2$ matrices are shown in table I.
In \cite{NPAMS} two of us used Fubini?s theorem to show that all relevant integrals converge and integrations can be interchanged. The interpretation of the transform used as isometric representations of the corresponding complex extension then generates the bi-orthonormal partner of an orthonormal bases and the arguments of section two ensures convergence of all integerals without fort her proof.
\begin{table}

 \centering
  \begin{tabular}{llll}
   \hline
  TF &Kernel & Exponential  & Exponential\\
  &&growth over $q^\prime$   & growth over $q$\\  \hline
MB  & $ e^{-\alpha(q^\prime -q)^2} $& $[-\beta,2];\alpha>\beta>0 $ & $[-\alpha\left(\frac{\alpha}{\alpha-\beta}-1\right),2]$ \\
H & $ e^{-\frac12\alpha {q^\prime}^2+2iq^\prime q+(1-\alpha)q^2} $& $[-\beta,2];\frac12\alpha<\beta<\infty$ & $[-\frac12\alpha\frac{\beta+\alpha}{\beta-\alpha},2]$ \\
SW &  $e^{-\alpha(q^\prime -iq)^2} $& $[-\beta,2];\alpha<\beta<\infty$ & $[-\alpha\left(1+\frac{\alpha}{\beta-\alpha}-1\right),2]$ \\ \hline
\end{tabular}
\caption{For three transforms (TF $\rightarrow $ MB: Margenau-Brink \cite{BM}, H: Hackenbroich \cite{HHH} and SW: S${\rm \ddot u}$nkel - Wildernuth \cite{SW}) we show the kernels as well as ranges for the exponential growth of the basis functions that may readily be mapped; The notation $[\beta, \nu]$ indicates an asymptotic behavior as $\exp(\beta|r|^\nu)$ for large $r$.}
 \end{table}

Actually we can go one step further: It is quite usual in these nuclear problems to strongly restrict the Hilbert space in some of the variables either to a single function or a very small finite subspace. For example the space on which the internal variables of an $\alpha-cluster$ live is often limited to a single Gaussian, a few Gaussians or some other finite set of functions. In this case we end up with a partial isometry, as a subspace of one of the dual pair of spaces is mapped too a subspace of the other. Thus the methods used in this context of nuclear physics can be considered in a unified way using the approach presented in section 2.

A wider range of applications has appeared in classical optics, where the paraxial approximation reduces the ray optics problem to a Hamiltonian one but the introduction of the waves leads to the use of representations of canonical transformations by Gaussian kernels, some of which correspond to complex extensions. 
The ensuing argumentation is similar to the one for the nuclear physics case, but extensive literature based on \cite{loebl} is available \cite{Wolfbook} with some more recent work e.g. refs \cite{CCTnum, CCTQopt,CCTsemiC} to which we wish to refer the reader. 

\section{Bosonic and fermionic operator Fock spaces}

We now describe in the above formal setting the procedure 
which has been termed `third quantization'; namely of constructing the Fock 
spaces of operators in which Liouvillian dynamics of open many-body quantum systems can be treated in a canonical way.

Let us assume we have $m$ bosonic degrees of freedom described by operators $a_j,j=1,\ldots,m$, and $n$ fermionic degrees 
of freedom described by operators $c_k,k=1,\ldots n$, which satisfy the following graded algebra:
\begin{eqnarray}
[a_j, a_k] &=& 0, \quad [a_j, a^*_k] = \delta_{j,k}, \\
\{ c_j, c_k \} &=& 0,\quad \{ c_j,  c^*_k \} = \delta_{j,k}, \quad [a_j, c_k] = 0, \quad  [a_j, c^*_k] = 0, 
\end{eqnarray}
where $[a,b]=a b - b a$, $\{a,b\}=a b + b a$ and $a^*_j$, $c^*_j$ are the Hermitian adjoint operators (the so-called creation 
operators) defined with respect to the standard Fock space representation.
Namely, taking the Fock vacuum $\psi_0$, a vector which is characterized by the property
\begin{equation}
a_j\psi_0  = 0,\quad c_k \psi_0 = 0,
\end{equation}
the complete countable basis of Fock  space $F$ can be constructed as
\begin{equation}
\psi_{\underline{i}} = \frac{(a_1^*)^{i_1}}{\sqrt{i_1!}}\frac{(a_2^*)^{i_2}}{\sqrt{i_2!}}\cdots \frac{(a_m^*)^{i_m}}{\sqrt{i_m!}}(c_1^*)^{i_{m+1}}(c_2^*)^{i_{m+2}}\cdots (c_n^*)^{i_{m+n}} \psi_0,
\end{equation}
where $\underline{i} = (i_1,\ldots i_{m+n})$ is the multi-index with the first $m$ components running over $\ZZ_+$ and the last $n$ components running over $\ZZ_2$.
\footnote{In physics usually a concrete function representation is chosen with a natural inner product in $F$, such that it becomes a Hilbert space and $\psi_{\underline{i}}$ forms an orthonormal basis.}

In treating Liouvillean dynamics in open quantum systems  \cite{prosen1, prosen2, prosenseligman1} one has to study physical states which are trace-class 
elements of ${\rm End}(F)$.
Taking any linear operator $f$ over $F$, we define two linear operators over ${\rm End}(F)$ as left and right multiplication maps
\begin{equation}
\hat{f}^{\rm L}x = f x,\qquad \hat{f}^{\rm R}x = x f.
\end{equation}
Let us now define a set of $4(n+m)$ maps - linear operators over $V$.
\begin{eqnarray}
\hat{a}_{0,j} &=& \hat{a}^{\rm L}_j ,\,\;\qquad   \hat{a}'_{0,j} =  \hat{a^*}^{\rm L}_j , \cr
\hat{a}_{1,j} &=& \hat{a^*}^{\rm R}_j ,\qquad   \hat{a}'_{1,j} =  \hat{a}^{\rm R}_j , \cr
\hat{c}_{0,j} &=& \hat{c}^{\rm L}_j ,\;\;\;\quad\quad   \hat{c}'_{0,j} =  \hat{c^*}^{\rm L}_j , \cr
\hat{c}_{1,j} &=& \hat{c^*}^{\rm R}_j \hat{\cal P} ,\;\;\quad   \hat{c}'_{1,j} =  \hat{c}^{\rm R}_j \hat{\cal P}.
\label{eq:maps}
\end{eqnarray}
The parity map $\hat{\cal P}$ is uniquely defined by requiring that it fixes two particular elements of  ${\rm End}(F)$, namely the unit operator $\one$ 
and the orthogonal projector to vacuum 
$\psi_0$, call it $v_0$, and that it commutes/anticommutes with bosonic/fermionic maps
\begin{eqnarray}
&&\hat{\cal P}\one = \one,\qquad
\hat{\cal P}v_0 = v_0,\cr
&&[\hat{\cal P},\hat{a}_{\nu,j}] = [\hat{\cal P},\hat{a}'_{\nu,j}] = \{\hat{\cal P},\hat{c}_{\nu,j} \} = \{\hat{\cal P},\hat{c}'_{\nu,j}\}=0.
\end{eqnarray}
A particular realization of $\hat{\cal P}$ is, for example, given in terms of the Wigner-Jordan phase operator:
\begin{equation}
\hat{\cal P} = \hat{\eta}^{\rm L} \hat{\eta}^{\rm R},\quad \eta = \exp\left(-\ii\pi \sum_{j=1}^n c^*_j c_j\right).
\end{equation}
One can straightforwardly check that the maps $\hat{a}_{\nu,j},\hat{c}_{\nu,j}$ again satisfy the canonical Bose-Fermi graded algebra:
\begin{eqnarray}
[\hat{a}_{\nu,j}, \hat{a}_{\mu,k}] &=& 0, \quad [\hat{a}_{\nu,j}, \hat{a}'_{\mu,k}] = \delta_{\nu,\mu}\delta_{j,k}, \\
\{ \hat{c}_{\nu,j}, \hat{c}_{\mu,k} \} &=& 0,\quad \{ \hat{c}_{\nu,j},  \hat{c}'_{\mu,k} \} = \delta_{j,k}, \quad [\hat{a}_{\nu,j}, \hat{c}_{\mu,k}] = 0, \quad  [\hat{a}_{\nu,j}, \hat{c}'_{\mu,k}] = 0, 
\end{eqnarray}
where $\mu,\nu\in\ZZ_2$.

We then formulate a sequence of elements of $V$ as
\begin{equation}
v_{\underline{i}} =  \prod_{\nu}\prod_{j=1}^m \frac{(\hat{a}'_{\nu,j})^{i_{\nu,j}}}{\sqrt{i_{\nu,j}!}} \prod_{\nu}\prod_{k=1}^n (\hat{c}'_{\nu,k})^{i_{\nu,m+k}} v_0.
\label{vi}
\end{equation}
with a $2(m+n)$ component multiindex $\underline{i}= (i_{\nu,j};\nu\in\{0,1\},j=1\ldots m+n)$.
Sequence $v_{\underline{i}}$ spans a Fock-space $V \subset {\rm End}(F)$ of Hilbert-Schmidt operators with a vacuum vector $v_0$.
It contains all physical density-state operators $\rho=\rho^*$, since they obey $\tr \rho^2 < \infty$, but it does not 
contain important physical observables, such as for example boson numbers $a^*_j a_j$.

However, we define the {\em relevant} Fock-Hilbert space of unbounded physical operators $Y$ by the $\ell^2$ closure of the following operator sequence:
\begin{equation}
y_{\underline{i}} =  \prod_{\nu}\prod_{j=1}^m \frac{(\hat{a}'_{\nu,j})^{i_{\nu,j}}}{\sqrt{i_{\nu,j}!}} \prod_{\nu}\prod_{k=1}^n (\hat{c}'_{\nu,k})^{i_{\nu,m+k}} \one.
\label{wi}
\end{equation}
The sequences $v_{\underline{i}}$ and $y_{\underline{i}}$ now define the product $\times$ in $V\times Y$ and the scalar products $\cdot$ 
within each space, according to which the vectors (\ref{vi}) and (\ref{wi}) form orthonormal and  bi-orthonormal sequences. The maps $\hat{a}_{\nu,j}$, $\hat{c}_{\nu,j}$ 
are defined in either of the space and by definition of the adjoint map 
$\hat{f}^*$, $x \cdot \hat{f} y = \hat{f}^*  x \cdot y$, one finds that
\begin{equation}
\hat{a}^*_{\nu,j} = \hat{a}'_{\nu,j},\quad \hat{c}^*_{\nu,j} = \hat{c}'_{\nu,j}.
\end{equation}

As a specific application, one may consider the equation of motion for the open Fermi-Bose system which is given by the Lindblad equation
\begin{equation}
\frac{\dd\rho }{\dd t} = \LL\rho :=
-\ii [H,\rho] + \sum_{m} \left(2 L_m \rho L_m^* - \{L_m^* L_m,\rho\} \right)
\label{eq:lind}
\end{equation}
where $H$ is a Hermitian operator (Hamiltonian) and $L_m$ are arbitrary (Lindblad) operators representing couplings to different
baths. 
The technique described above gives us a straightforward procedure of writing the generator (Liouvillian) $\LL$ in terms of canonical maps $\hat{a}_{\nu,j}$, $\hat{c}_{\nu,j}$, once the Hamiltonian and the Lindblad operators $L_m$ are given in terms of $a_j$ and $c_j$.
Indeed:
\begin{equation}
\LL = -\ii \hat{H}^{\rm L} + \ii \hat{H}^{\rm R} + \sum_m (2 \hat{L}_m^{\rm L} \hat{L^*}_m^{\rm R} - 
\hat{L^*}_m^{\rm L} \hat{L}_m^{\rm L} - 
\hat{L^*}_m^{\rm R} \hat{L}_m^{\rm R}) 
\end{equation}
which by identifications (\ref{eq:maps}) can be transformed to a desired form. For example, quadratic Hamiltonians $H$ with linear Lindbladians $L_m$ 
map to quadratic Liouvilleans $\LL$.
In order for a phase map $\hat{\cal P}$ to cancel out, the expression for the Hamiltonian has to be even (say, quadratic) in fermionic operators, 
as $\hat{\cal P}^2=1$, as in the dissipators it always cancels out due to combination of $L_m$ and $L_m^*$ in all the terms.

For example, in physics the typical Fermi-Bose term in the Hamiltonian, which represents interaction between fermions mediated by a boson, is written as
\begin{equation}
T=c^*_j c_k a_l +{\rm h.c.}
\end{equation}
In third quantization this then maps to
\begin{equation}
\hat{\cal T} = \hat{T}^{\rm L} -  \hat{T}^{\rm R} = \hat{c}^*_{0,j} \hat{c}_{0,k} \hat{a}_{0,l} +  \hat{c}_{1,j} \hat{c}^*_{1,k} \hat{a}^*_{1,l} +{\rm h.c.}
\end{equation}

\section{Conclusions and outlook}
We have seen, that a bilinear form between elements of two vector spaces, such as e.g. the trace of a product of two matrices allows to establish measures on both
 spaces that allow to identify them as finite or infinite dimensional Hilbert spaces. This construction, albeit almost trivial, solves the problem related to the 
space of operators that includes the relevant observables, which typically are not trace-class. 
While the operative result is the same as in \cite{prosenseligman2}, the structure conjectured in that paper is now demonstrated to be valid. 
Furthermore it is clear, that this simple framework will apply to any equivalent formulation that does not explicitly 
invoke Fock space, simply because the trace of the product of a state with an observable is a central tenet of the theory of open quantum systems, and that alone is
 sufficient to perform the construction. Note though, that the measure on the two Hilbert spaces depends on the explicit bilinear form 
 and on the vectors included in the space. If we make the very reasonable assumption that the identity forms part of the observable 
 space then it follows immediately that all states are trace-class matrices as we would probably like them to be. A short comment on the 
meaning of quantities like purity or entropy of a mixed quantum state seems relevant as they take us outside the framework we introduced. 
Yet if we keep in mind that they do not correspond to observables in the sense discussed here, we have to ask how we can measure them. 
In the case of purity there  is an elegant prescription in terms of a simultaneous measurement on two states \cite{puritymeasurement} 
and an inspection of this technique shows, that it brings us back to the present framework, 
using states of the doubled system and appropriate observables.

The possibilities of alternate applications are tempting. An obvious one would be to exploit the relation of the Bargmann transform to the generating function of 
Hermite polynomials and use generating functions for other functions in an analogous way. For example, as Bargmann himself pointed out and Barut and Girardello \cite{BarutG} 
explained in detail, a similar construction can be found for radial oscillator functions. Using standard generating functions and our formal procedure without choosing optimal complex domains, might lead to ugly expressions, but avoiding their explicit use, most problems can be formulated in terms of the simple bilinear form. 

Summarizing, we may say that we present a very simple construction, which allows to present rather widespread applications of isometries and partial isometries both old and very recent under a unified scheme. This scheme is also promising to simplify further development in problems that may arise with a similar mathematical structure of dual spaces or dual bases

We acknowledge financial support by the program P1-0044 of the Slovenian Research Agency and by Deutsche Forschungsgemeinschaft  in the SPP 15-16 
as well as by  UNAM/DGAPA/PAPIIT, grant IG10113 and CONACyT, grant 154586.
One of us (T.H.S.) is acknowledging a Fellowship for distinguished scholars by the Slovenian Research Agency.

\end{document}